\documentclass[aps,floats,prl,
twocolumn]{revtex4-1}

\usepackage{amsfonts,amsmath} \usepackage{bm} \usepackage{dcolumn}
\usepackage{epsfig} \usepackage{latexsym}

\begin{document}

\title{Many-body eigenstate thermalization from one-body quantum chaos: emergent arrow of time}
%\title{Ehrenfest time, many-body eigenstate thermalization and emergent arrow of time}
%\title{Eigenstate thermalization without interaction and genuine randomness}
%\title{From eigenstate typicality to thermalization without interaction: role of ergodicity}

\author{Chushun Tian$^1$, Kun Yang$^{2,3}$, and Jiao Wang$^4$}

\affiliation{$^1$Institute for Advanced Study, Tsinghua University, Beijing 100084, China\\
$^2$National High Magnetic Field Laboratory and Department of Physics,
Florida State University, Tallahassee, Florida 32306, USA\\
$^3$Department of Physics, Tsinghua University, Beijing 100084, China\\
$^4$3Department of Physics and Institute of Theoretical Physics and Astrophysics, Xiamen University, Xiamen, Fujian 361005, China}

%\date{\today}
\begin{abstract}

A profound quest of statistical mechanics is the origin of irreversibility -- the arrow of time \cite{Boltzmann1872,von_Neumann29,Bogoliubov46,Prigogine62,Zaslavsky81,Krylov79,gogolinreview,Gogolin15,rigolreview,Izrailev16}. New stimulants have been provided, thanks to unprecedented degree of control reached in experiments
with isolated quantum systems \cite{gogolinreview,Gogolin15,rigolreview,Izrailev16,Reimann16} and rapid theoretical developments \cite{Huse14} of many-body localization in disordered interacting systems \cite{Altshuler06}. The
proposal of (many-body) eigenstate thermalization (ET) \cite{Deutsch91,Srednicki94,Rigol08} for these systems reinforces the common belief that either interaction or extrinsic randomness
is required for thermalization.
%\cite{gogolinreview,rigolreview,Deutsch91,Srednicki94,Rigol08,Izrailev16,Reimann16}.
Here, we unveil a quantum thermalization mechanism challenging this belief. We find that, provided one-body quantum chaos is present, as a pure many-body state evolves the arrow of time can emerge, even without interaction or randomness.
%, and further be reversed. Specifically, in
In times much larger than
the Ehrenfest time \cite{Larkin68} that signals the breakdown of quantum-classical correspondence, quantum chaotic motion leads to thermal -- Fermi-Dirac (FD) or Bose-Einstein (BE) -- distributions and thermodynamics in individual eigenstates.
%; this process is reversed after long times.
Our findings lay dynamical foundation of statistical mechanics and thermodynamics of isolated quantum systems.

\end{abstract}

\maketitle

Isolated quantum systems, ranging from the universe to complex nuclei, exhibit a wealth of physical phenomena. For such systems the standard ensemble description
%, often justified by subjecting the system to external randomness,
is not applicable. Can statistical mechanics and thermodynamics emerge from unitary (pure state) evolution then? How does irreversibility reconcile with this reversible evolution? Studies of these
problems were initiated by von Neumann soon after the birth of quantum mechanics \cite{von_Neumann29}, and
%owing to realizations of a variety of isolated quantum systems
have become urgently important recently \cite{gogolinreview,Gogolin15,rigolreview,Izrailev16}.
%, partially due to experimental realizations of various isolated quantum systems.
The ET hypothesis \cite{Deutsch91,Srednicki94,Rigol08} -- that the expectation value of a physical observable in an
eigenstate equals to its statistical mechanics ensemble average -- lies at the heart of the renewal of interests in this old subject. The concept of ET is central to many-body localization \cite{Altshuler06} currently under intense investigations \cite{Huse14}.

\begin{figure}
\includegraphics[width=8.7cm] {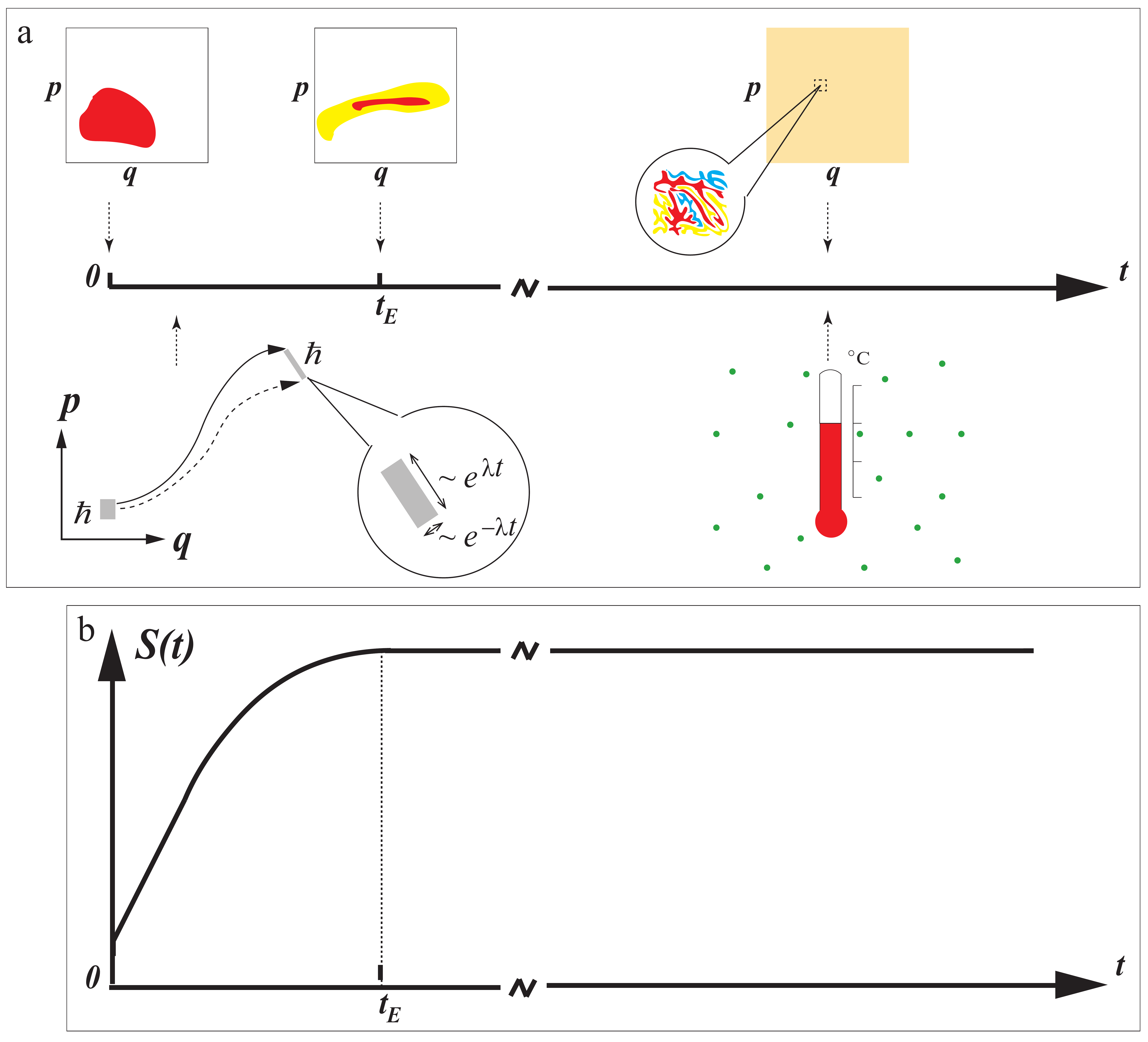}
\caption{{\bf Emergent arrow of time. a}, as time passes a Planck's cell is deformed due to instability of single-particle classical motion (below $t$-axis, left). This causes correlation functions (in the Wigner representation, with the value represented by different colors)
to follow Liouville evolution up to
%the Ehrenfest time
$t_E$,
%which signals the breakdown of quantum-classical correspondence,
and exhibit quantum structures in later times (above $t$-axis).
Associated with the formation of these structures,
an ideal gas (green solid circles) approaches ET, with thermodynamics established for individual eigenstates (below $t$-axis, right).
{\bf b}, the time profile of entropy (\ref{eq:S45}) corresponds to the many-body pure state evolution in {\bf a}.
%associated with single-particle quantum motion.
}
\label{fig:3}
\end{figure}

This hypothesis has received increasingly strong support from numerics \cite{Rigol08,rigolreview}.
It stands to reason that it could significantly improve understandings of irreversibility, namely, the arrow of time. To pave the way to such advance a number of conceptual difficulties, however, have to be overcome. First of all, this hypothesis is mute to evolution, because it addresses the long-time average of observables, but not observables at long times. Most importantly,
it determines neither under what circumstances ET occurs, nor whether genuine thermal equilibrium can be achieved, even if ET occurs. In fact, whether and how thermodynamics of entire isolated
system (not subsystem as addressed in most
literatures) emerges from unitary evolution is a largely unexplored realm.
%In addition, the formulation of this hypothesis suffers certain
%conceptual ambiguities, notably, the arbitrariness in the choice of physical observables with their expectation values expected to be thermal: there are
%important observables
%,%(hermitian operators), especially those involving many degrees of freedom
%which do {\em not} respect ET hypothesis.
%A prominent example is $-\rho\ln\rho$ with $\rho$ being the system's density matrix:
%Naively, this is expected to correspond to thermal or von Neumann entropy which is non-zero in a statistical mechanics description; however,
%in a pure state it has a vanishing expectation value and thus cannot be used to probe thermalization.
%This issue becomes prominent as numerical experiments are concerned.
A common belief, conforming to the canonical paradigm in statistical mechanics \cite{Prigogine62}, is that particle interaction \cite{Srednicki94} or randomness \cite{Deutsch91}, managing to redistribute particle energies, is an essential ingredient giving rise to ET. In particular, in Ref.~\cite{Srednicki94} it is shown analytically how interaction leads to thermal (Maxwell-Boltzmann, FD and BE) distributions for individual eigenstates, and the result was numerically found \cite{Casati98} to hold even for interacting systems as small as including only two particles.

This common belief has been challenged by the eigenstate typicality (ETp) very recently found for a
remarkably simple system -- a free Fermi gas put on a torus \cite{Yang15}.
It was shown that, in a typical highly excited eigenstate, the reduced density matrix of subsystem may approach
that of certain thermal ensemble.
%, provided the size of the subsystem is much smaller than that of the whole system.
(Such a property
%was later termed ``canonical typicality of energy eigenstates'' \cite{NoteonETp}, which
resembles, but is stronger than the canonical typicality \cite{Lebowitz06,Popescu06}.)
%Among the key consequences of
%ETp is that the entanglement entropy between the subsystem and its complement is equal to the thermal entropy of the subsystem in an appropriate ensemble.
%Of conceptual importance of the discovery of ETp suggests is
So, even without interaction or randomness, the subsystem can be thermalized. This finding provides new insights to quantum thermalization in isolated systems and many questions arise. Notably, in isolated systems without interaction or randomness, can ET, which is much stronger than ETp (in the sense that
ETp guarantees that all local
%hermitian operators -- with support in the subsystem --
physical observables have thermal expectation values, while
%, and as such escapes the aforementioned conceptual ambiguity of ET.
ET requires not only local, but also non-local
%physical
observables, to be thermal), be achieved? In such simple systems, can irreversibility and thermodynamics emerge from reversible evolution of pure states?

Here we study these problems analytically and unveil a new scenario of quantum thermalization. The scenario requires, counter-intuitively, neither interaction nor randomness, rather, the indistinguishability of identical particles -- an intrinsic quantum many-body effect -- and one-body quantum chaos. The latter has two-fold meanings. First, in the classical limit the single-particle
motion is dynamically unstable or more precisely {\it mixing} \cite{Zaslavsky81}, which leads to ergodicity and exponential decay of correlations in time.
As shown in Fig.~\ref{fig:3}a (below $t$-axis, left),
a small deviation in initial conditions (e.g., velocity and position)
is exponentially amplified in later times with a rate -- the Lyapunov exponent $\lambda$. In the 1940s, Krylov demonstrated the importance of mixing to classical statistical mechanics, and showed that $\lambda$ is order of the inverse scattering time \cite{Krylov79}. Second, this chaoticity is {\it not} suppressed by quantum interference.
(The quantization of a classical chaotic system is not necessarily chaotic \cite{note_2}.)
%; equivalently, the single-particle spectral fluctuations obeys the Wigner-Dyson statistics \cite{Bohigas84}.

Figure \ref{fig:3}a is a schematic representation of the new scenario of quantum thermalization. It shows that the two ingredients above, hand in hand, lead to thermalization at times much larger than the Ehrenfest time \cite{Larkin68},
\begin{equation}\label{eq:19}
    t_E\sim \lambda^{-1}\ln (A/\hbar).
\end{equation}
Here $\hbar$ is the Planck's constant and $A$ a characteristic classical action, and we pay no attention to the overall numerical coefficient. Physically, $t_E$ arises from dynamical instability
(below $t$-axis, left). Indeed, the Planck's cell in phase space
-- a result of quantum uncertainty -- is deformed in the course of time.
At $t_E$ this cell is macroscopically large in certain directions and a classical trajectory is thereby split.
So, $t_E$ is a hallmark of the breakdown of quantum-classical correspondence.
We probe thermalization via physical observables whose expectation values can be expressed
in terms of correlation functions. Due to dynamical instability the correlation function follows
the classical evolution in short times and relaxes to a thermal equilibrium value at times $t\gg t_E$.
%(i$\rightarrow$iii).
That value is found to follow the ET hypothesis. It is independent of the initial state as well as details of microscopic dynamics. Moreover, it is essentially an average with respect to the FD or BE distribution, implying that genuine thermal equilibrium (not microcanonical distribution as studied in most literatures \cite{Rigol08}) is established (below $t$-axis, right). So, an arrow of time -- the evolution towards ET and thermodynamics -- is seen to emerge from reversible, deterministic evolution of many-body pure states. Surprisingly, neither interaction nor randomness is required. [We ignore quantum recurrence since this occurs at large times and practically is irrelevant, see Supplementary Material (SM) for discussions.] Importantly, we find that $t_E$ is much larger than the time scale for thermalization \cite{Tasaki15} obtained based on typicality. Our findings
lay dynamical foundation of thermalization of isolated quantum systems.

\begin{figure}
\includegraphics[width=8.7cm] {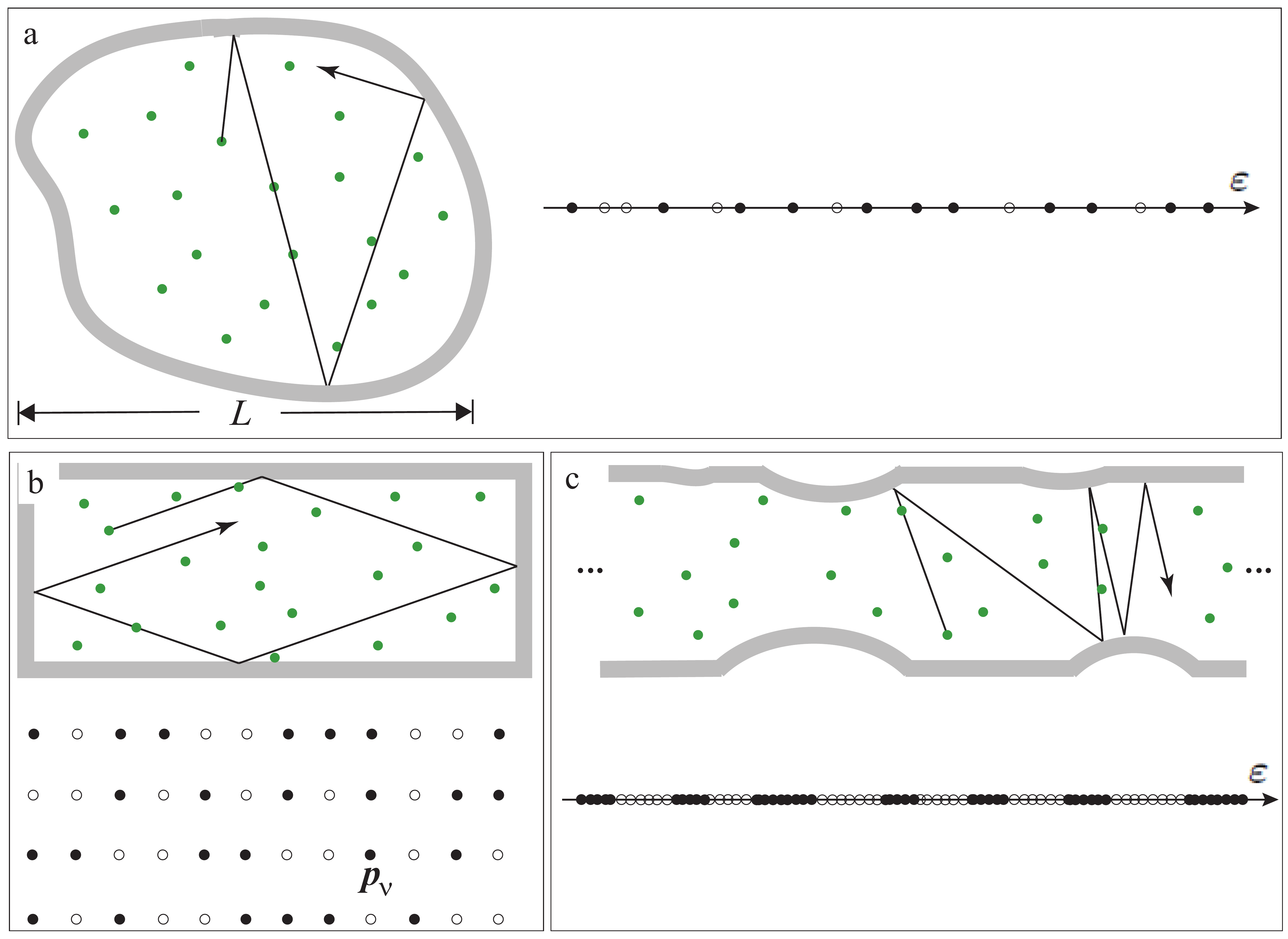}
\caption{{\bf Ideal gas confined in cavities.} {\bf a}, a chaotic cavity (left) confining ideal gas (green solid circles) and giving rise to
%ergodic single-particle quantum motion
one-body quantum chaos. For ensuing single-particle quantum motion energy is the only good quantum number, with a discrete, chaotic spectrum (right).
%{\bf b $\&$ c}, cavities
%supporting non-ergodic quantum motion.
%where one-body quantum chaos does not arise.
{\bf b}, for rectangular cavity
%(left),
the single-particle quantum motion is completely integrable, and the discrete momenta $\boldsymbol{p}_\nu$ consist of a complete set of good quantum numbers.
{\bf c}, for an infinitely long quasi-one-dimensional cavity with rough surface,
%(right),
the motion is non-ergodic due to Anderson localization, and a dense point spectrum follows, with energy as the only good quantum number again. For Fermi gas, a typical excited eigenstate corresponds to a random occupation configuration in good quantum number space. The black solid (empty) circles implies that a single-particle eigenstate is (un)occupied.}
\label{fig:1}
\end{figure}

{\it Setup and analytical results.} We put large but finite $N$ identical particles free of interaction in a $d$-dimensional cavity of volume $V$
and size $L$ (Fig.~\ref{fig:1}). The particle undergoes specular reflection on the boundary. For a cavity where one-body quantum chaos arises (Fig.~\ref{fig:1}a, left), the single-particle eigenenergy $\varepsilon$ is the only good quantum number. The corresponding spectrum exhibits chaotic fluctuations (Fig.~\ref{fig:1}a, right), with a
%mean level spacing $\Delta$ and a
spectral density $\rho(\varepsilon)$. The eigenstate of an ideal gas confined in the cavity corresponds to a configuration of occupation number $\boldsymbol{m}\equiv\{n_\nu\}$, where $\nu$ labels single-particle eigenstate.

We prepare a
%\cite{note_BEC}
pure state $F$ which is highly excited so that the
expectation value of energy per particle, $E/N$, is much larger than the mean level spacing.
%the BE condensation is excluded.
This state is superposed by eigenstates $\boldsymbol{m}$, the corresponding eigenenergies of which are located in a narrow energy shell, with a width much smaller (larger) than $E$ (mean level spacing). It evolves, $F\rightarrow F(t)$, following the Schr{\"o}dinger equation. Noting that a one-body physical observable can be expressed in terms of the correlation function, $M_{\boldsymbol{r}\boldsymbol{r}'}(t)\equiv\langle F(t)| a^\dagger_{\boldsymbol{r}'}a_{\boldsymbol{r}}|F(t)\rangle$, between the spatial points $\boldsymbol{r}$ and $\boldsymbol{r}'$ at time $t$, with $a_{\boldsymbol{r}}$ ($a^\dagger_{\boldsymbol{r}}$) being the annihilation (creation)
operator at $\boldsymbol{r}$, we focus on this function below.

We find analytically that $\overline{M_{\boldsymbol{r}\boldsymbol{r}'}(t)}$ relaxes to a genuine thermal value.
Here the overline stands for the average with respect to $\boldsymbol{q}\equiv \frac{\boldsymbol{r}+\boldsymbol{r}'}{2}$
over several de Broglie wavelengths. This wavelength corresponds to the average single-particle energy $E/N$ and is much smaller than $L$.
The average accounts for finite resolution in realistic measurements. More precisely,
%for $t_E\ll t\ll t_{rec.}-t_E$,
\begin{equation}\label{eq:1}
    \overline{M_{\boldsymbol{r}\boldsymbol{r}'}(t)}\stackrel{t\gg t_E}{\rightarrow}\frac{1}{V}\int d\varepsilon \rho(\varepsilon)f\left(\frac{|\boldsymbol{r}-\boldsymbol{r}'|}{\lambda_\varepsilon}\right)\frac{1}{e^{\frac{\varepsilon-\mu}{T}}\pm1}.
\end{equation}
The second factor in the integrand is given by
\begin{equation}\label{eq:13}
    f(x)=\Gamma(d/2)(x/2)^{-\frac{d-2}{2}}J_{\frac{d-2}{2}}(x),\,\,\, x=|\boldsymbol{r}-\boldsymbol{r}'|/\lambda_\varepsilon,
\end{equation}
where $\Gamma(x)$ is the gamma function and $J_\nu (x)$ the Bessel function, and
%. Equation (\ref{eq:13}) shows that this factor oscillates in $|\boldsymbol{r}-\boldsymbol{r}'|$, with a period comparable to
the de Broglie wavelength $\lambda_\varepsilon\equiv \frac{\hbar}{\sqrt{2m\varepsilon}}$ ($m$ the particle mass). The last factor in (\ref{eq:1}) is the FD (BE) distribution, where $T$ is the temperature (the Boltzmann constant set to unity) and $\mu$ the chemical potential. We further show that these two emergent quantities, $T$ and $\mu$, and the fundamental thermodynamic relation,
\begin{equation}\label{eq:20}
    TdS=dE-\mu dN,
\end{equation}
with $S$ being the emergent thermal entropy, all hold for individual eigenstates of ideal gas.
%Note that in the FD (BE) distribution $\varepsilon$ is the single-particle eigenenergy, rather than the usual kinetic energy.
Finally, $\lambda \sim \sqrt{\frac{E/N}{mL^2}}\sim \sqrt{\frac{T}{mL^2}}$ and $A\sim \sqrt{\frac{mEL^2}{N}}\sim \sqrt{mTL^2}$. Combining with Eq.~(\ref{eq:19}) they give $t_E\gg\frac{\hbar}{T}$, where $\frac{\hbar}{T}$ is the thermalization time scale obtained based on typicality \cite{Tasaki15}.

Equation (\ref{eq:1}) has transparent physical meanings. The left-hand side respects faithfully
reversible, deterministic, unitary
evolution; the right-hand side reflects ET and thermal equilibrium. They are connected via the arrow of time, and $t\gg t_E$ indicates that thermalization is a result of the formation of phase space quantum structures (Fig.~\ref{fig:3}a). We further find an entropy $S(t)$ characterizing the emerging of the arrow of time (Fig.~\ref{fig:3}b), which coincides with the thermal entropy in equilibrium.

In the absence of one-body quantum chaos, the results summarized above do not hold: neither does the arrow of time emerge, nor the thermal distribution. In Fig.~\ref{fig:1} we give two examples. One is an ideal gas confined in a rectangular cavity (b) and the other in an infinitely long quasi-one-dimensional cavity with rough surface (c). The single-particle quantum motion in the former cavity is completely integrable, while in the latter exhibits Anderson localization.

{\it Analytic theory.}
%To study the evolution of correlation function
%part and the other induced by quantum impurities which exists inevitably in real systems. The latter part is much weaker than the former and vanishes in the classical limit.
We pass to the Wigner representation,
\begin{eqnarray}
%\overline{
M_{\boldsymbol{r}\boldsymbol{r}'}(t)
%}
\equiv\int d\boldsymbol{p} e^{-\frac{i}{\hbar}(\boldsymbol{r}-\boldsymbol{r}')\cdot \boldsymbol{p}}
%\overline{
M(\boldsymbol{q},\boldsymbol{p};t)
%}
.
\label{eq:15}
\end{eqnarray}
%where $M(\boldsymbol{q},\boldsymbol{p};t)$ is the Wigner transformation of $M_{\boldsymbol{r}\boldsymbol{r}'}(t)$.
In SM we show that the Wigner function defined above, $M(\boldsymbol{q},\boldsymbol{p};t)$, obeys
\begin{equation}\label{eq:14}
    \left(\partial_t-\{H(\boldsymbol{q},\boldsymbol{p}),\,\cdot\,\}_{{\rm Moyal}}
    %-D\frac{\partial^2}{\partial\boldsymbol{p}^2}
    \right)M(\boldsymbol{q},\boldsymbol{p};t)=0.
\end{equation}
Here $H(\boldsymbol{q},\boldsymbol{p})=\frac{\boldsymbol{p}^2}{2m}+V(\boldsymbol{q})$ is the classical Hamiltonian, with $V(\boldsymbol{q})$ being the potential. The Moyal bracket \cite{Moyal49} $\{H,\,\cdot\,\}_{{\rm Moyal}}\equiv -\frac{i}{\hbar}\sin(i\hbar\{H,\,\cdot\,\})$, where the Poisson bracket $\{H,\,\cdot\,\}\equiv (\partial_{\boldsymbol{q}}H \cdot \partial_{\boldsymbol{p}}-\partial_{\boldsymbol{p}}H\cdot \partial_{\boldsymbol{q}})(\cdot)$.
To make the Moyal bracket well defined we have assumed that $V(\boldsymbol{q})$ is analytic.
This assumption is technical and inessential to the physical results summarized above. This evolution is of one-body nature, with many-body aspects of $M$ entering into the initial condition $M(\boldsymbol{q},\boldsymbol{p};0)$.
%, and as shown in SM exhibits recurrent (quasiperiodic) behavior.
Moreover, Eq.~(\ref{eq:14}) is reversible, being invariant under the
time reversal: $\boldsymbol{q}\rightarrow \boldsymbol{q}, \boldsymbol{p}\rightarrow -\boldsymbol{p}, t\rightarrow -t$, and deterministic. In the
%The momentum diffusion operator arises from quantum impurities, and the diffusion coefficient $D$ vanishes in the classical limit.
limit $\hbar\rightarrow 0$, the Moyal bracket reduces to the Poisson bracket and Eq.~(\ref{eq:14}) to the Liouville equation.

We recall a fundamental property \cite{Zaslavsky81} of chaotic classical motion (Fig.~\ref{fig:3}a, below $t$-axis, left). Due to dynamical instability
%of classical motion
a volume element expands exponentially in unstable directions with a rate $\lambda$,
%. Since the volume is conserved, this volume element
and shrinks exponentially in stable directions with the same rate for compensation.
Upon deforming and folding (due to the boundedness of phase space), the volume element develops
fine structures at a scale
%momentum
$\sim
%\sigma_p
e^{-\lambda t}$ which decays exponentially in time.
%, where $\sigma_p$ is the initial size of the patch in momentum direction.
This classical dynamical property has far-reaching consequences on the quantum evolution (\ref{eq:14}) (Fig.~\ref{fig:3}a, above $t$-axis).
Observing the structure of Moyal bracket, $\{H,\,\cdot\,\}_{{\rm Moyal}}=\{H,\,\cdot\,\}+\sum_{n=1}^\infty {\cal O}(\hbar^{2n}\partial_{\boldsymbol{q}}^{2n+1}V\partial_{\boldsymbol{p}}^{2n+1})$, we find that, due to the shrinking of phase space volume in stable directions, the value of higher order derivative terms of $\{H,\,\cdot\,\}_{{\rm Moyal}}$ increase exponentially in time, i.e., ${\cal O}(\hbar^{2n}\partial_{\boldsymbol{q}}^{2n+1}V\partial_{\boldsymbol{p}}^{2n+1})
=(\frac{\hbar e^{\lambda t}}{A})^{2n}{\cal O}(\partial_{\boldsymbol{q}}V\partial_{\boldsymbol{p}})
%={\cal O}(\hbar^{2n} \partial_{\boldsymbol{q}}^{2n+1}V e^{(2n+1)\lambda t}/\sigma_p^{2n+1})
$. These terms become comparable to the Poisson bracket at $t_E$.
%,
%with $A\gg \hbar$ being a characteristic action,
%\begin{equation}\label{eq:17}
%    t_E=\frac{1}{\lambda}\ln \frac{\chi \sigma_p}{\hbar}.
%\end{equation}
%when the quantum-classical correspondence breaks down.
As the time further increases, $M(\boldsymbol{q},\boldsymbol{p};t)$ oscillates more and more rapidly in phase space coordinates, and the oscillation structure saturates eventually. When the time is reversed, the expansion and shrinking processes are exchanged as a result of
reversibility, and fine structures are formed again in the course of time.
%: this occurs to the backward evolution from $t_{rec.}$ (v$\rightarrow$iii).
% how the arrow of time emerges
%and is reversed in the course of time.

The observable $\overline{M_{\boldsymbol{r}\boldsymbol{r}'}(t)}$ is governed by the coarse graining of $M(\boldsymbol{q},\boldsymbol{p};t)$. Indeed, the spatial average over de Broglie oscillations introduces coarse graining of $\boldsymbol{q}$-coordinate. Besides, the exponential in Eq.~(\ref{eq:15}) serves as a ``low-pass filter'': fine structures of $M(\boldsymbol{q},\boldsymbol{p};t)$ developed in momentum scales $\ll \hbar/|\boldsymbol{r}-\boldsymbol{r}'|$ do not contribute to the momentum integral. In other words, this exponential introduces coarse graining of $\boldsymbol{p}$-coordinate. Taking these into account, we can reduce Eq.~(\ref{eq:15}) to
\begin{equation}\label{eq:16}
    \overline{M_{\boldsymbol{r}\boldsymbol{r}'}(t)}=\int d\boldsymbol{p} e^{-\frac{i}{\hbar}(\boldsymbol{r}-\boldsymbol{r}')\cdot \boldsymbol{p}} M_{c.g.}(\boldsymbol{q},\boldsymbol{p};t),
\end{equation}
with $M_{c.g.}(\boldsymbol{q},\boldsymbol{p};t)$ being the coarse graining of $M(\boldsymbol{q},\boldsymbol{p};t)$. We emphasize that the coarse graining, whose details are irrelevant and will be exemplified
in SM, does {\it not} violate the unitarity of evolution. In SM we further show
%$M_{c.g.}$ follows the Liouville evolution, i.e., $\partial_tM_{c.g.}=\{H,\,M_{c.g.}\}$ for $t\gg t_E$. Because $M_{c.g.}$ is sufficiently smooth in phase space, following the well-known mathematical theorems in the ergodic theory $M_{c.g.}(\boldsymbol{q},\boldsymbol{p};t\rightarrow \infty)$ must be uniform in the phase space energy shell. This gives
\begin{eqnarray}\label{eq:17}
    M_{c.g.}(\boldsymbol{q},\boldsymbol{p};t) \stackrel{t\gg t_E}{\rightarrow} {\rm function\, of}\, H(\boldsymbol{q},\boldsymbol{p}).
    %\nonumber\\
    %{\rm for}\, t_E\ll t\ll t_{rec.}-t_E \quad \quad
\end{eqnarray}
It suggests that when quantum-classical correspondence breaks down $M(\boldsymbol{q},\boldsymbol{p};t)$ is uniform in the phase space energy shell upon coarse graining. We also simulate quantum evolution (\ref{eq:14}) for a simplified system and confirm this result (see SM). The result (\ref{eq:17}) is by no means classical. On the contrary, it arises from that all quantum corrections cancel out and thereby is highly nonperturbative [in ($\{H,\,\cdot\,\}_{{\rm Moyal}}-\{H,\,\cdot\,\}$)]. A similar phenomenon was found previously in a study of Ehrenfest time effects on level statistics of systems with unitary symmetry \cite{Larkin04}.

Equations (\ref{eq:16}) and (\ref{eq:17}) show that $\overline{M_{\boldsymbol{r}\boldsymbol{r}'}(t)}$ relaxes
%a thermal (at this moment the temperature has not yet emerged and by ``thermal'' we do not mean genuine thermal equilibrium.)
at $t\gg t_E$ and stays at equilibrium value in later times. (Recall that throughout we do not pay attention to quantum recurrence occurring at extremely long times.)
Importantly, this result is regarding $\overline{M_{\boldsymbol{r}\boldsymbol{r}'}(t)}$ at long times,
rather than the long time average -- as studied in most literatures of quantum thermalization \cite{Deutsch91,Srednicki94,Rigol08} --
of $\overline{M_{\boldsymbol{r}\boldsymbol{r}'}(t)}$! As
shown in SM, Eq.~(\ref{eq:17}) has
a connection to the origin of classical irreversibility namely the Ruelle-Pollicott resonance \cite{Ruelle87,Pollicott85} in chaotic dynamics.
%A result similar to (\ref{eq:17}) was obtained previously in studies of quantum-classical correspondence \cite{Prozen06}.

Because $\overline{M_{\boldsymbol{r}\boldsymbol{r}'}(t)}$ is at equilibrium in long times, the equilibrium value is $\lim_{T\rightarrow\infty} \int_0^T \frac{dt}{T} \overline{M_{\boldsymbol{r}\boldsymbol{r}'}(t)}$.
To calculate this time average we substitute $|F\rangle=\sum_{\boldsymbol{m}} C_{\boldsymbol{m}} |\boldsymbol{m}\rangle$ (with $\sum_{\boldsymbol{m}}|C_{\boldsymbol{m}}|^2=1$), into the definition of
$M_{\boldsymbol{r}\boldsymbol{r}'}(t)$,
%we can reduce this time average to
%$\overline{\langle \boldsymbol{m}| a^\dagger_{\boldsymbol{r}'}a_{\boldsymbol{r}}|\boldsymbol{m}\rangle}$,
%where $\boldsymbol{m}$ is a typical highly excited eigenstate composing $F$ and a small correction arising from atypical eigenstates has been ignored. This gives
obtaining
\begin{equation}\label{eq:18}
    \overline{M_{\boldsymbol{r}\boldsymbol{r}'}(t)}\stackrel{t\gg t_E}{\rightarrow}\sum_{\boldsymbol{m}}|C_{\boldsymbol{m}}|^2\overline{\langle \boldsymbol{m}| a^\dagger_{\boldsymbol{r}'}a_{\boldsymbol{r}}|\boldsymbol{m}\rangle}.
\end{equation}
Equations (\ref{eq:16})-(\ref{eq:18}) merely show that $\overline{M_{\boldsymbol{r}\boldsymbol{r}'}(t)}$ relaxes when the system evolves
%forwards or backwards
for sufficiently long times ($\gg t_E$). They do not indicate whether and
to what extent the system can be thermalized.
%Can the ET hypothesis \cite{Deutsch91,Srednicki94,Rigol08} be applicable to the right-hand side of Eq.~(\ref{eq:18})?
%he relaxation of the observable $\overline{M_{\boldsymbol{r}\boldsymbol{r}'}}$ leads to the realizing of ET.
%We emphasize that this does not necessarily imply the validity of thermodynamics:
It is precisely at this point that the principle of indistinguishability of identical particles comes to play key roles. In the remainder we will show explicitly and analytically how this principle and Eq.~(\ref{eq:18}), hand in hand, lead to ET and the thermal (FD or BE) distribution, justifying that thermodynamic notions such as the temperature and Boltzmann entropy are emergent objects of unitary evolution.

To calculate the right-hand side of (\ref{eq:18}) we start from an exact expression,
$\langle \boldsymbol{m}| a^\dagger_{\boldsymbol{r}'}a_{\boldsymbol{r}}|\boldsymbol{m}\rangle=\sum_\nu n_\nu C_\nu(\boldsymbol{r}-\boldsymbol{r}',\boldsymbol{q})$,
where $C_\nu(\boldsymbol{r}-\boldsymbol{r}',\boldsymbol{q})\equiv \psi_\nu(\boldsymbol{r})\psi^*_\nu(\boldsymbol{r}')$ is
the autocorrelation of single-particle eigenfunction $\psi_\nu$.
%\begin{equation}\label{eq:3}
%    M_{\boldsymbol{r}\boldsymbol{r}'}=\sum_\nu n_\nu \psi_\nu(\boldsymbol{r})\psi^*_\nu(\boldsymbol{r}')
%    \equiv \sum_\nu n_\nu C_\nu(\boldsymbol{r}-\boldsymbol{r}',\boldsymbol{q}),
%\end{equation}
%for the matrix elements of the correlation matrix, where
%$\psi_\nu$ is the single-particle eigenfunction with the normalization $\int d\boldsymbol{r} |\psi_\nu(\boldsymbol{r})|^2=1$.
%Equation (\ref{eq:3}) trades the correlation matrix to the autocorrelation,
%$C_\nu(\boldsymbol{r}-\boldsymbol{r}',\boldsymbol{q})$, of single-particle eigenfunction.
%For a highly excited eigenstate of fermi gas,
%the occupied single-particle eigenstates ($n_\nu=1$)
%have a characteristic single-particle eigenenergy $\varepsilon$, which is so large that the corresponding de Broglie wavelength
%\begin{equation}\label{eq:2}
%    \lambda_\varepsilon\ll L,
%\end{equation}
%with $L$ being the characteristic length of the cavity.
%This semiclassical condition makes it legitimate \cite{Berry77} to perform for $C_\nu(\boldsymbol{r}-\boldsymbol{r}',\boldsymbol{q})$. This gives
%\begin{eqnarray}
%C_\nu(\boldsymbol{r}-\boldsymbol{r}',\boldsymbol{q})&\approx& \overline{C_\nu(\boldsymbol{r}-\boldsymbol{r}',\boldsymbol{q})}\nonumber\\
%&=& \int d\boldsymbol{p} e^{\frac{i}{\hbar}(\boldsymbol{r}-\boldsymbol{r}')\cdot \boldsymbol{p}} \overline{\Psi_{\nu}(\boldsymbol{q},\boldsymbol{p})},
%\label{eq:4}
%\end{eqnarray}
%where $\Psi_\nu$ is the Wigner transformation \cite{Gutzwiller90} of $\psi_\nu$.
Following Berry \cite{Berry77}, in the presence of one-body quantum chaos, the Wigner transformation of $\psi_\nu(\boldsymbol{r})$, denoted as $\Psi_{\nu}(\boldsymbol{q},\boldsymbol{p})$, is given by
\begin{equation}\label{eq:5}
    \overline{\Psi_{\nu}(\boldsymbol{q},\boldsymbol{p})}=\frac{\delta(\varepsilon-H(\boldsymbol{q},\boldsymbol{p}))}{\int\!\!\int d\boldsymbol{q}d\boldsymbol{p}\delta(\varepsilon-H(\boldsymbol{q},\boldsymbol{p}))}.
\end{equation}
Note that we have performed the averaging with respect to $\boldsymbol{q}$ over several de Broglie wavelengths, and the microcanonical distribution (\ref{eq:5}) is defined on the single(!)-particle phase space. Physically, this distribution is a manifestation of ergodicity of single-particle
motion.
%which is fundamentally different from previous work \cite{Srednicki94}
%where the many-body version of (\ref{eq:5}) is employed to formulate ET hypothesis for interacting many-body systems.
There is a correction to the distribution (\ref{eq:5}), which arises from the quantum scar \cite{Heller84}. Since $\psi_\nu$ corresponds to single-particle eigenenergy $\varepsilon$ much larger than the mean level spacing, effects of this correction on $C_\nu(\boldsymbol{r}-\boldsymbol{r}',\boldsymbol{q})$ are negligible (see SM). Using Eq.~(\ref{eq:5}), we find
\begin{eqnarray}
\overline{C_\nu(\boldsymbol{r}-\boldsymbol{r}',\boldsymbol{q})}=\frac{\int d\boldsymbol{\Omega}
e^{-\frac{i}{\lambda_\varepsilon}\boldsymbol{\Omega}\cdot(\boldsymbol{r}-\boldsymbol{r}')}}{V\int d\boldsymbol{\Omega}}\equiv\frac{1}{V}f\left(\frac{|\boldsymbol{r}-\boldsymbol{r}'|}{\lambda_\varepsilon}\right)\!\!.
\,\,\,\,\,\,\,\,\,\label{eq:6}
%\\
%f\left(\frac{|\boldsymbol{r}-\boldsymbol{r}'|}{\lambda_\varepsilon}\right)
%&=&\Gamma\left(d/2\right)\frac{J_{\frac{d-2}{2}}\left(\frac{|\boldsymbol{r}-\boldsymbol{r}'|}{\lambda_\varepsilon}\right)}{\left(\frac{|\boldsymbol{r}-\boldsymbol{r}'|}{2\lambda_\varepsilon}\right)^{\frac{d-2}{2}}},\nonumber
\end{eqnarray}
Here $\boldsymbol{\Omega}$ is the solid
angle, and the integral over $\boldsymbol{\Omega}$ is calculated in SM giving Eq.~(\ref{eq:13}) \cite{Berry77}.
%Equation (\ref{eq:6}) obeys the general rules (\ref{eq:16}) and (\ref{eq:17}).
$f(x)$ has the same form for different single-particle eigenstates satisfying $\lambda_\varepsilon \ll L$. It can be shown (see SM) that
for two nearest
%single-particle eigenenergies
$\varepsilon,\varepsilon'$,
\begin{equation}\label{eq:8}
    f\left(\frac{|\boldsymbol{r}-\boldsymbol{r}'|}{\lambda_\varepsilon}\right)\approx f\left(\frac{|\boldsymbol{r}-\boldsymbol{r}'|}{\lambda_{\varepsilon'}}\right)
\end{equation}
because of one-body quantum chaos (Fig.~\ref{fig:2}a). As shown below, combining with the indistinguishability of identical particles, it gives rise to ET and thermal distributions.

In general, most eigenstates contributing to the initial state are typical \cite{Yang15}. Each of such eigenstates corresponds to a random configuration $\{n_\nu\}$ (cf. Fig.~\ref{fig:1}). Thanks to Eq.~(\ref{eq:8}), we can follow Ref.~\cite{Yang15} and introduce the coarse graining of the occupied single-particle eigenenergy space. Specifically, we divide it into a number of clusters (Fig.~\ref{fig:2}a),
each of which, labeled by $m$, has a length $\delta \varepsilon_m$ and includes $g_m
%\approx \frac{\delta \varepsilon_m}{\Delta_\varepsilon}
\gg 1$ single-particle eigenenergies. The $m$th cluster is occupied by $N_m\gg 1$ particles. The configuration $\{N_m\}$ defines a macrostate, constrained by
$\sum_m N_m=N$ and $\sum_m N_m \varepsilon_m=E$,
%\begin{equation}\label{eq:10}
%    \sum_m N_m=N,\quad \sum_m N_m \varepsilon_m=E,
%\end{equation}
where $\varepsilon_m$ is the center energy of the $m$th cluster, and the eigenenergy of ideal gas can be approximated by $E$. Taking these and Eqs.~(\ref{eq:6}) and (\ref{eq:8}) into account, we obtain
\begin{eqnarray}\label{eq:9}
    \overline{\langle \boldsymbol{m}| a^\dagger_{\boldsymbol{r}'}a_{\boldsymbol{r}}|\boldsymbol{m}\rangle}
    %&=&
    %\overline{\sum_m N_m C_m(\boldsymbol{r}-\boldsymbol{r}',\boldsymbol{q})}\nonumber\\&=&
    =\frac{1}{V}\sum_m g_m n(\varepsilon_m) f\left(\frac{|\boldsymbol{r}-\boldsymbol{r}'|}{\lambda_{\varepsilon_m}}\right),
\end{eqnarray}
where
%$C_m$ is the autocorrelation of single-particle eigenfunction corresponding to the single-particle eigenenergy $\varepsilon_m$, and
$n(\varepsilon)$ varying smoothly in $\varepsilon$ (Fig.~\ref{fig:2}b) is the occupation number averaged locally over a large enough single-particle eigenenergy window. Equation (\ref{eq:9}) establishes the relation between the equilibrium value of correlation function and a macrostate. From this we see that many microscopic states lead to the same equilibrium correlation function, and the typicality implies the proliferation of eigenstates with the same macroscopic features.

\begin{figure}
\includegraphics[width=8.7cm] {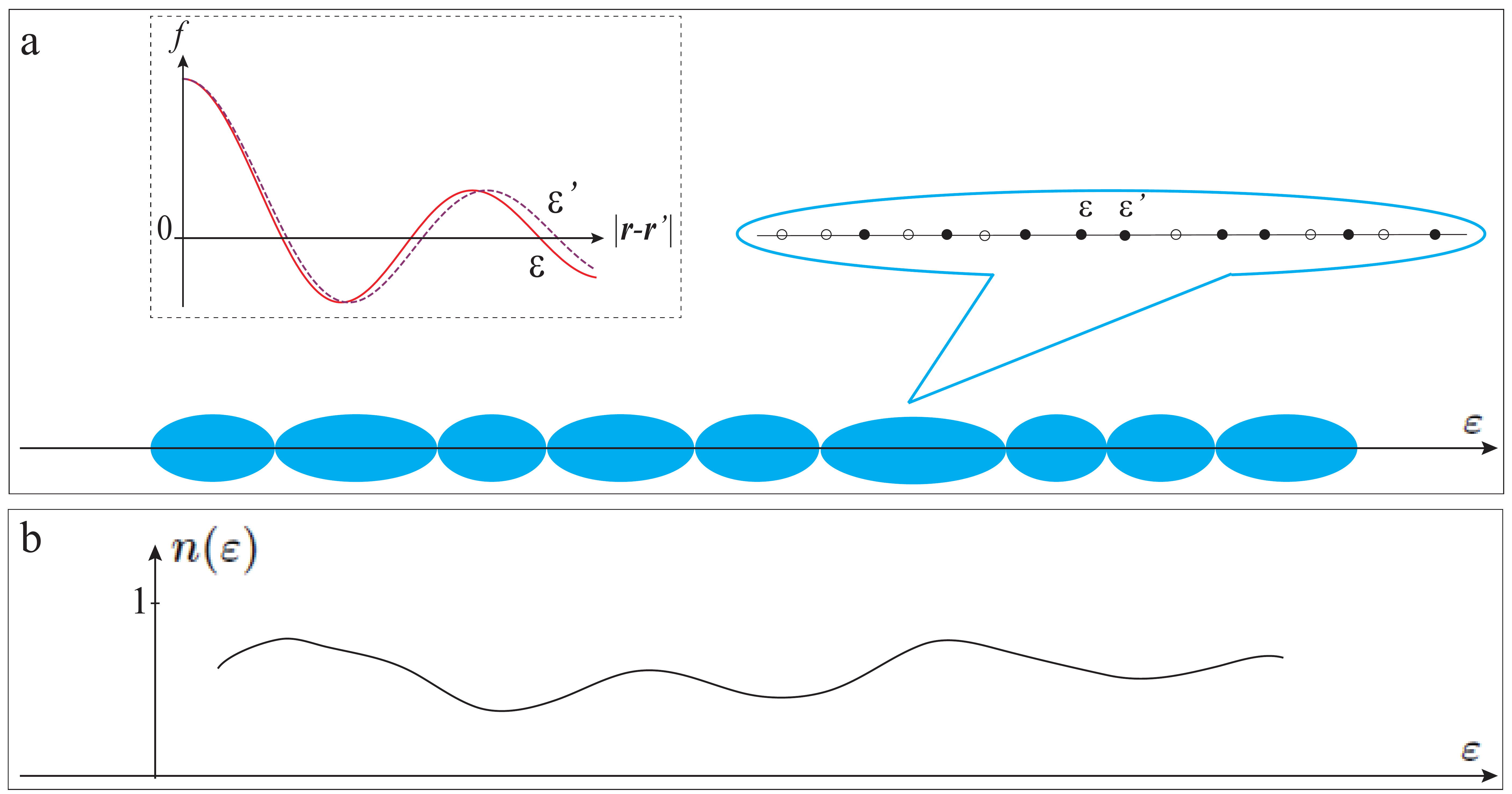}
\caption{{\bf Coarse graining of single-particle eigenenergy space.} {\bf a}, a macrostate is an occupation configuration associated with the coarse grained single-particle eigenenergy space (blue clusters). This coarse graining is allowed because, given $\boldsymbol{r}-\boldsymbol{r}'$, the values of $f$ are closed to each other for two nearest eigenenergies $\varepsilon,\varepsilon'$ (inset). {\bf b}, this macrostate is described by a smooth function $n(\varepsilon)$, namely, the occupation number averaged over the local single-particle eigenenergy spectrum.}
\label{fig:2}
\end{figure}

We now decide the configuration $\{N_m\}$ in Eq.~(\ref{eq:9}). From the typicality we expect it to be most probable, i.e.,
%corresponds to a largest number of microstates implying
$\delta[\ln W(\{N_m\})-\sum_m
(\alpha +\beta \varepsilon_m)N_m]=0$.
%\begin{eqnarray}\label{eq:11}
%    \delta\left[\ln W(\{N_m\})+\sum_m \left(\alpha +\beta \varepsilon_m\right)N_m\right]=0.
%\end{eqnarray}
Here, the number of microstates $W(\{N_m\})$ is $\prod_m\frac{g_m!}{N_m!(g_m-N_m)!}$ for Fermi gas and $\prod_m\frac{(N_m+g_m-1)!}{N_m!(g_m-1)!}$ for Bose gas. $\alpha,\beta$ are the Lagrange multipliers associated with the constraints
above.
%(\ref{eq:10}).
Then, the most probable $\{N_m\}$ corresponds to
\begin{equation}\label{eq:12}
    n(\varepsilon_m)=\frac{1}{e^{\beta \varepsilon_m+\alpha}\pm 1}.
\end{equation}
This gives the FD and BE distribution, respectively. The extremal value condition further gives $\frac{\partial S}{\partial E}=\beta\equiv \frac{1}{T}$ and $\frac{\partial S}{\partial N}=\alpha\equiv-\frac{\mu}{T}$, with the thermal (Boltzmann) entropy $S=\ln W$. This justifies the thermodynamic relation (\ref{eq:20}). We emphasize that Eq.~(\ref{eq:12}) results from the coarse grained single-particle eigenenergy space. If the condition (\ref{eq:8}) is not satisfied, this space does not exist and genuine thermal equilibrium (\ref{eq:12}) does not follow (see below for further discussions). This is very different from standard derivations of the FD and BE distribution \cite{Landau58}, which are irrespective of system's dynamical properties.

Because the energy window is much narrower than $E$, the fluctuations in the (many-body) eigenenergy can be ignored. Combined with the above constraints
on total particle number and eigenenergy
%(\ref{eq:10})
this implies that, for different $\boldsymbol{m}$, the Lagrange multipliers take the same value, and so does $n(\varepsilon_m)$. Following Eq.~(\ref{eq:9}) we find that the values of $\overline{\langle \boldsymbol{m}| a^\dagger_{\boldsymbol{r}'}a_{\boldsymbol{r}}|\boldsymbol{m}\rangle}$ are the same for different typical eigenstates $\boldsymbol{m}$. This justifies the ET hypothesis made in Refs.~\cite{Deutsch91,Srednicki94} and confirmed numerically in Ref.~\cite{Rigol08}. Additionally, Eq.~(\ref{eq:20}) shows that equilibrium thermodynamics holds even for individual eigenstates. Furthermore, Eq.~(\ref{eq:18}) reduces to $\overline{M_{\boldsymbol{r}\boldsymbol{r}'}(t)}\rightarrow \overline{\langle \boldsymbol{m}| a^\dagger_{\boldsymbol{r}'}a_{\boldsymbol{r}}|\boldsymbol{m}\rangle}$ for $t\gg t_E$, with a negligible correction arising from atypical eigenstates. Therefore, the equilibrium value of $\overline{M_{\boldsymbol{r}\boldsymbol{r}'}(t)}$ is independent of initial states, i.e., $C_{\boldsymbol{m}}$.

We substitute Eq.~(\ref{eq:12}) as well as $g_m \approx\delta \varepsilon_m\rho(\varepsilon_m)$ into Eq.~(\ref{eq:9}). Passing to the continuum limit, we reduce Eq.~(\ref{eq:9}) to the right-hand side of Eq.~(\ref{eq:1}). In combination with Eq.~(\ref{eq:18}), we justify Eq.~(\ref{eq:1}) completely.

Finally, we define an entropy [$\tilde n\equiv (2\pi\hbar)^d M_{c.g.}({\boldsymbol{q}},{\boldsymbol{p}};t)$],
\begin{eqnarray}\label{eq:S45}
    S(t)
    %&
    \equiv
    %&
    -\int \!\!\!\!\int\!\! \frac{d\boldsymbol{q} d\boldsymbol{p}}{(2\pi\hbar)^d} \big(|\tilde n
    %_{c.g.}(
    %{\boldsymbol{q}},{\boldsymbol{p}};t)
    |\ln|\tilde n
    %_{c.g.}(
    %{\boldsymbol{q}},{\boldsymbol{p}};t)
    |
    %\nonumber\\&&
    \pm|1\mp\tilde n
    %_{c.g.}({\boldsymbol{q}},{\boldsymbol{p}};t)
    |\ln|1\mp\tilde n
    %_{c.g.}({\boldsymbol{q}},{\boldsymbol{p}};t)
    |\big).
\end{eqnarray}
In SM we show that its evolution can be represented schematically by Fig.~\ref{fig:3}b. The increasing
%(decreasing)
process of $S(t)$ for $t\lesssim t_E$
%($t_{rec.}-t_E\lesssim t \leq t_{rec.}$)
corresponds to the emerging
%(reversing)
of the arrow of time. In equilibrium ($t\gg t_E
%\ll t \ll t_{rec.}-t_E
$), $S(t)$ coincides with the thermal entropy $S$. As predicted by von Neumann \cite{von_Neumann29} and numerically observed in Ref.~\cite{Wang01}, even in equilibrium stage, entropy can be suppressed from its thermal value from time to time (see SM for discussions). This may be attributed to quantum structures of certain eigenstates.

{\it Absence of one-body quantum chaos.} Our theory requires finite $t_E$ (i.e., $\lambda>0$) and the existence of $n(\varepsilon)$. None of them is present in the absence of one-body quantum chaos, which can be caused by two mechanisms as follows.

(i) the single-particle motion is non-chaotic, i.e., $\lambda=0$, already at the classical level (Fig.~\ref{fig:1}b). Inheriting from classical Liouville integrability the single-particle quantum motion is completely integrable. First, because of $t_E\rightarrow \infty$ the expectation value of observable does not relax. Second, for this single-particle motion a good quantum number, i.e., the discrete momentum $\boldsymbol{p}_\nu$, results. Correspondingly, $C_\nu(\boldsymbol{r}-\boldsymbol{r}',\boldsymbol{q})\sim e^{i\boldsymbol{p}_\nu\cdot (\boldsymbol{r}-\boldsymbol{r}')}$. Because the distance between two nearest $\boldsymbol{p}_\nu$ is ${\cal O}(1/L)$, and $\boldsymbol{r}-\boldsymbol{r}'={\cal O}(L)$, for two nearest eigenstates the phase (i.e., the exponent of $C_\nu$) difference is ${\cal O}(1)$. This leads to a significant change in the value of $C_\nu$, invalidating the coarse graining of single-particle eigenenergy space in Fig.~\ref{fig:2} and making the function $n(\varepsilon)$ ill defined.

(ii) classical chaos exists, i.e., $\lambda>0$, but is suppressed by quantum interference,
as exemplified by motion in a sufficiently long quasi-one-dimensional cavity with rough surface (Fig.~\ref{fig:1}c). For single-particle motion in this cavity it has been known \cite{Larkin96} that at $t\gtrsim t_E$ the crossover from diffusion to Anderson localization
%\cite{Gertsenshtein59}
takes place. For the latter the single-particle eigenenergies constitute a dense point spectrum and the eigenfunction exhibits exponential localization in the longitudinal direction. For two eigenenergies infinitesimally closed to each other, the eigenfunctions are well separated in space. As a result, the eigenfunction correlation between two given points ($\boldsymbol{r},\boldsymbol{r}'$) changes substantially as the eigenenergy varies slightly. Therefore, the single-particle eigenenergy space cannot be coarse grained, and
$n(\varepsilon)$ is again ill defined.
%So, ET does not occur. Indeed, for $t\gg t_E$ the (static) conductivity is fully suppressed: no dissipation -- required for thermalization -- is generated.

{\it Discussion.} We have seen that the roles of one-body quantum chaos are two-fold. First, it gives rise to a coarse grained single-particle
eigenenergy space or more precisely Eq.~(\ref{eq:9}). This coarse graining, together with the indistinguishability of identical particles, leads to ET with the FD (BE) distribution as a key manifestation. Second, it leads to the emerging
%and reversing
of the arrow of time from unitary evolution. Strikingly, neither interaction nor randomness is required. Our theory is established for one-body observables. Whether it can be generalized to many-body observables remains unclear.

Our findings are not restricted to ideal gas. They also provide new perspectives of quantum thermalization of isolated interacting systems. For example, upon applying the standard mean field approximation the single-particle orbital often emerges from interacting systems. This orbital -- an effective single-particle quantum motion -- is chaotic, provided the interaction is strong and the energy of orbital is high. In this case we expect our theory to be applicable. A realistic system composed of strongly interacting electrons is the gold ion Au$^{25+}$, where the emergence of the FD distribution has been seen numerically near the ionization threshold \cite{Flambaum99}.

We would like to thank H.-H. Lai and J.-B. Gong for discussions. This work is supported by the National Natural Science Foundation of China (Grant No. 11535011) and the National Science Foundation (Grants No. DMR-1004545 and No. DMR-1442366).

\end{document}